\documentclass[11pt,twoside]{article}
\usepackage[cp1251]{inputenc}
\usepackage{amssymb,amsmath,amscd}
\usepackage{fancyheadings}
\usepackage[russian]{babel}
\usepackage{bm}
\usepackage{graphicx}
\makeatother
\usepackage{wrapfig}
\newcommand{\GOD}{2023}
\newcommand{\UDK}[1]{\noindent{\footnotesize\sl УДК #1}}
\newcommand{\Nazva}[1]{\begin{center}\baselineskip=6.0mm{\Large\textbf{#1}}\end{center}\vspace*{0.5mm}}
\newcommand{\Avtor}[1]{\centerline{\large\textbf{\copyright~\GOD~г. \ #1}}\vspace*{-4.0mm}}
\newcommand{\AVTOR}{~}
\newcommand{\NAZVA}{~}
\newcommand{\lit}[3]{\vspace*{0.7mm}\par\noindent\makebox[5.2mm][r]{#1.~}\parbox[t]{159.8mm}{{\textit{#2}}~{#3}}\hspace*{-1.6mm}}
\begin{document}

\renewcommand{\abstractname}{}

\centerline{\large\textbf{УРАВНЕНИЯ С ЧАСТНЫМИ ПРОИЗВОДНЫМИ}}

%
%
%

\vspace{2mm}
\hrule
\vspace{2mm}

\UDK{517.925}
\Nazva{РЕШЕНИЯ АНАЛОГОВ ВРЕМЕННЫХ УРАВНЕНИЙ ШРЕДИНГЕРА, СООТВЕТСТВУЮЩИХ
ПАРЕ ГАМИЛЬТОНОВЫХ СИСТЕМ $H^{2+2+1}$}
\Avtor{В.~А.~Павленко}
\renewcommand{\NAZVA}{РЕШЕНИЯ АНАЛОГОВ ВРЕМЕННЫХ УРАВНЕНИЙ ШРЕДИНГЕРА}
\renewcommand{\AVTOR}{Павленко}

\thispagestyle{empty}

\begin{abstract}\noindent
Настоящая работа продолжает серию работ, в котрых строятся $2\times2$ матричные совместные решения двух скалярных эволюционных уравнений, 
которые являются аналогами временных уравнений Шредингера. Эти уравнения соответствуют гамильтоновой системе 
$H^{2+2+1}$, являющейся одним из представителей иерархии вырождений изомонодромной системы Гарнье. 
Упомянутую иерархию описал Х. Кимура в 1986 году. В терминах решений линейных систем обыкновенных 
дифференциальных уравнений метода изомнодромных деформаций, условием совместности которых являются 
гамильтоновы уравнения системы $H^{2+2+1}$, конструируемые совместные матричные решения аналогов временных уравнений 
уравнений Шредингера в настоящей работе будут выписаны явно.
\medskip\\
\end{abstract}

\bigskip

\textbf{Введение.}
Б. И. Сулейманов в 1988 г. в работах~[1, 2] построил решения линейных эволюционных уравнений вида 
\begin{equation}\label{PavlenkoVA1}
\frac{\partial}{\partial t }\Psi=H(t,x,-\frac{\partial}{\partial x})\Psi.
\end{equation}
В уравнениях \eqref{PavlenkoVA1} из работ~[1, 2] линейные дифференциальные операторы $H(t,x,-\frac{\partial}{\partial x})$  
соответствуют квадратичным по импульсам $p$ гамильтонианам $H=H(t,q,p)$ гамильтоновых систем 
\begin{equation}\label{PavlenkoVA2}
q'_t=H'_p(t,q,p),\qquad p'_t=-H'_q(t,q,p),
\end{equation}

\textbf{Замечание 1.} {\it Эти системы таковы, что если в \eqref{PavlenkoVA2} исключить $p$, то получится одно из шести 
обыкновенных дифференциальных уравнений (ОДУ) Пенлеве. Уравнения \eqref{PavlenkoVA1} получаются из соответствующих временных уравнений Шредингера
\begin{equation*}
i\hbar \frac{\partial}{\partial t}\Psi=
H(t,x,-i\hbar\frac{\partial}{\partial x})\Psi,
\end{equation*} 
зависящих от постоянной Планка $h=2\pi\hbar$, в результате формальной подстановки $\hbar=-i$. }

В терминах решений линейных систем методом изомонодромных деформаций (ИДМ), которые выписаны в статье 
Р. Гарнье~[3], решения \eqref{PavlenkoVA1} в работах~[1, 2] предъявлены явно. При этом условием совместности 
упомянутых линейных систем являются шесть соответствующих классических ОДУ Пенлеве.

Известно, что все эти ОДУ могут быть получены из шестого уравнения при помощи 
процедур последовательного вырождения. Другими словами уравнения Пенлеве представляют собой некоторую 
иерархию, которую можно изобразить в виде диаграммы:
\begin{center}
\begin{picture}(200,30)
\put(0,15){$H^6$}
\put(18,18){\vector(1,0){20}}
\put(40,15){$H^5$}
\put(55,15){\vector(2,-1){30}}
\put(55,15){\vector(2,1){30}}
\put(85,-5){$H^3$}
\put(85,25){$H^4$}
\put(102,30){\vector(2,-1){24}}
\put(102,0){\vector(3,2){24}}
\put(128,15){$H^2$}
\put(144,18){\vector(1,0){20}}
\put(167,15){$H^1$.}
\end{picture}
\end{center}
Здесь имеются ввиду упомянутые выше гамильтоновы системы $H_j$ $(j=1,\ldots,6)$ 
вида \eqref{PavlenkoVA2} для соответствующих уравнений Пенлеве. Каждая стрелка соответсвует процедуре вырождения одной из этих 
<<вышестоящих>> гамильтоновых систем к <<нижестоящей>>. 

Позже специфика связи решений уравнений ИДМ для уравнений Пенлеве с эволюционными 
уравнениями отмечалась и использовалась в следующих работах~[4 -- 15].

Помимо шести классических ОДУ Пенлеве в настоящий момент исследователи интересуются и другими нелинейными ОДУ 
более высокого порядка, которые также интегрируются ИДМ. На сегодня, в частности, известен (см.~[16 -- 20]) 
конечный список совместных пар гамильтоновых систем ОДУ
\begin{equation}\label{PavlenkoVA3}
(q_j)^\prime_{s_k}=(H_{s_k})^\prime_{p_j}, \quad 
(p_j)^\prime_{s_k}=-(H_{s_k})^\prime_{p_j}\quad (k=1,2)\quad (j=1,2)
\end{equation}
с гамильтонианами $H_{s_k}(s_1,s_2,q_1,q_2,p_1,p_2)$, каждое из которых есть условие совместности
двух линейных систем ОДУ вида 
\begin{equation}\label{PavlenkoVA4}
V^{\prime}_{s_k}=L_{s_k}V,
\end{equation}
\begin{equation}\label{PavlenkoVA5}
V^\prime_{\eta}=AV,
\end{equation} 
где квадратные матрицы $L_{s_k}$ и $A$ (матрица $A$ одна и та же для обеих гамильтоновых систем \eqref{PavlenkoVA3}) 
одинаковой размерности рациональны по переменной $\eta$. Соответствующие решения ОДУ, являющихся условием 
совместности таких пар, называются изомонодромными. К их числу относятся решения иерархии гамильтоновых вырождений 
системы Гарнье, выписанной в известной статье Х. Кимуры (см.~[16]). (Позднее Х. Кавамуко (см.~[20]) дополнил этот список). 

В работах~[16, 17] представлена аналогичная диаграмма вырождений для гамильтоновых систем иерархии Кимуры:

\begin{center}
\begin{picture}(400,50)
\put(0,25){$H^{1+1+1+1+1}$}
\put(60,28){\vector(1,0){30}}
\put(95,25){$H^{2+1+1+1}$}
\put(145,28){\vector(2,1){30}}
\put(145,22){\vector(2,-1){30}}
\put(180,40){$H^{2+2+1}$}
\put(180,5){$H^{3+1+1}$}
\put(220,45){\vector(1,0){30}}
\put(220,8){\vector(1,0){30}}
\put(220,40){\vector(4,-3){40}}
\put(220,12){\vector(4,3){40}}
\put(260,5){$H^{4+1}$}
\put(260,40){$H^{3+2}$}
\put(290,40){\vector(1,-1){10}}
\put(290,12){\vector(2,3){10}}
\put(300,25){$H^{5}$}
\put(315,28){\vector(1,0){30}}
\put(350,25){$H^{\frac 92}$.}
\end{picture}
\end{center}

\textbf{Замечание 2.} {\it Как сами уравнения Пенлеве, так и их высшие аналоги могут быть представлены через 
гамильтоновы системы с разными гамильтонианами, (см. монографию~[21] гл.2 раздел 2.8), а также работу~[22].}

Известно, что для всех представителей иерархии Кимуры справедливы две эквивалентные формы: форма 
совместных пар гамильтоновых систем \eqref{PavlenkoVA3}, определяемых квадратичными по импульсам $p_1$, $p_2$ и 
рациональными по координатам $q_1$, $q_2$ различными парами гамильтонианов $H_{s_k}(s_1,s_2,q_1,q_2,p_1,p_2)$, 
а также форма совместных пар гамильтоновых систем \eqref{PavlenkoVA3}, определяемых квадратичными 
по импульсам $p_1$, $p_2$ и полиномиальными по координатам $q_1$, $q_2$ 
различными парами гамильтонианов $H_{s_k}(s_1,s_2,q_1,q_2,p_1,p_2)$. 
Для большинства из этих гамильтоновых систем с двумя степенями свободы уже построены $2\times 2$ 
матричные совместные решения пар аналогов уравнений Шредингера 
\begin{equation}\label{PavlenkoVA6}
\varepsilon\Psi_{s_k}=H_{s_k}(s_1, s_2,-\varepsilon\frac{\partial}{\partial x}, 
-\varepsilon\frac{\partial}{\partial y},x,y)\Psi \quad (k=1,2)
\end{equation}
с $\varepsilon=1$,
соответствующие парам гамильтонианов $H_{s_k}(s_1,s_2,q_1,q_2,p_1,p_2)$ \linebreak $(k=1,2)$ этих изомонодромных систем. 
Решения эволюционных уравнений \eqref{PavlenkoVA6} для низших  представителей $H^{9/2}$ и $H^5$ вырождений системы Гарнье выписаны 
в работе Б.И. Сулеманова~[23]. Для первого представителя данной иерархии $H^{1+1+1+1+1}$ соответствующие решения 
представлены в работе Б.И. Сулейманова и Д.П. Новикова~[24]. Для вырождений $H^{2+1+1+1}$  и $H^{4+1}$ подобного рода
решения выписаны в работах автора совместно с Б.И. Сулеймановым, (см.~[25, 26] соответственно). Для еще 
одного вырождения, а именно, для $H^{3+2}$ автор представил соответствующие решения эволюционных уравнений в работе~[27].

В настоящей статье будут представлены $2\times 2$ матричные решения аналогов временных уравнений Шредингера c 
$\varepsilon=1$, которые соотвнтствуют гамильтоновой системе $H^{2+2+1}$. Эти решения будут представлены в 
двух формах: в рациональной и в полиномиальной. Другими словами, мы предъявим решения уравнений вида 
\begin{equation}\label{PavlenkoVA7}
\varepsilon\frac{\partial\Psi}{\partial\tau_j}=
H^{2+2+1}_{\tau_j}(\tau_1,\tau_2,x,y,-\varepsilon\frac{\partial}{\partial x},-\varepsilon\frac{\partial}{\partial y})\Psi
\quad (j=1,2), \quad (\varepsilon=1), 
\end{equation}
где дифференциальные операторы $H^{2+2+1}_{\tau_j}(\tau_1,\tau_2,x,y,-\varepsilon\frac{\partial}{\partial x},-\varepsilon\frac{\partial}{\partial y})$ 
соответствуют гамильтонианам с рациональными координатами. Затем мы предъявим решения уравненйи вида 
\begin{equation}\label{PavlenkoVA8}
\varepsilon\Psi_{s_k}=H^{2+2+1}_{s_k}(s_1, s_2,-\varepsilon\frac{\partial}{\partial r}, 
-\varepsilon\frac{\partial}{\partial \rho},r,\rho)\Phi \quad (k=1,2), \quad (\varepsilon=1),
\end{equation}
где дифференциальные операторы $H^{2+2+1}_{s_k}(s_1, s_2,-\varepsilon\frac{\partial}{\partial r},-\varepsilon\frac{\partial}{\partial \rho},r,\rho)$ 
соответствуют гамильтонианам с полиномиальными координатами. Соответствующие решения \eqref{PavlenkoVA7}, \eqref{PavlenkoVA8} явным 
образом будут выражены через совместные решения матричных линейных  пар  ИДМ \eqref{PavlenkoVA4}, \eqref{PavlenkoVA5} из 
статьи~[17], условием совместности которых являются гамильтоновы ОДУ \eqref{PavlenkoVA3}, соответствующие 
гамильтонианам системы $H^{2+2+1}$. 

Отметим, что решения уравнений типа временных уравнений Шредингера, которые будут строяться данной статье и те,  
которые были построены в ряде из упомянутых выше работ, представляют собой своеобразные специальные функции нового типа: 
несмотря на то, что они не могут быть выписаны в терминах интегралов типа Фурье--Лапласа, однако, в принципе, задача описания связи их поведения при  
\[
|s_1|+|s_2|+|x|+|y|\to \infty
\] 
в различных направлениях решается вполне эффективно даже для комплексных $s_1$, $s_2$, $x$, $y$. 
Более подробно по этому поводу см. работу Б.И. Сулейманова~[28, раздел 2.3].

\smallskip

\textbf{1. Различные формы системы $H^{2+2+1}$ и уравнения ИДМ для этой системы.} 
В статье~[16] гамильтонова система $H^{2+2+1}$ выписана в двух формах. В первой форме 
соответствующие гамильтонианы рациональны по координатам. Упомянутая система в этом случае имеет вид: 
\begin{equation}\label{PavlenkoVA9}
\frac{\partial\lambda_k}{\partial\tau_j}=\frac{\partial H_j}{\partial\mu_k},\quad 
\frac{\partial\mu_k}{\partial\tau_j}=-\frac{\partial H_j}{\partial\lambda_k}\quad (j,k=1,2)
\end{equation}
где гамильтонианы $H_i(\tau_1,\tau_2,\lambda_1,\lambda_2,\mu_1,\mu_2)$ задаются формулами: 
\begin{multline}\label{PavlenkoVA10}
\tau_1H_1=-\frac{\lambda_1^2(\lambda_1-1)^2(\lambda_2-1)}{\lambda_1-\lambda_2}\mu_1^2+
\frac{\lambda_2^2(\lambda_1-1)(\lambda_2-1)^2}{\lambda_1-\lambda_2}\mu_2^2+\\+
\frac{\lambda_1^2(\lambda_1-1)^2(\lambda_2-1)}{\lambda_1-\lambda_2}\left(\frac{\kappa_0}{\lambda_1}-
\frac{\gamma_1\tau_2}{\lambda_1^2}+\frac{\kappa_1-1}{\lambda_1-1}-\frac{\gamma_2\tau_1}{(\lambda_1-1)^2}\right)\mu_1-\\-
\frac{\lambda_2^2(\lambda_1-1)(\lambda_2-1)^2}{\lambda_1-\lambda_2}\left(\frac{\kappa_0}{\lambda_2}-
\frac{\gamma_1\tau_2}{\lambda_2^2}+\frac{\kappa_1-1}{\lambda_2-1}-\frac{\gamma_2\tau_1}{(\lambda_2-1)^2}\right)\mu_2-
\kappa(\lambda_1-1)(\lambda_2-1),
\end{multline}
\begin{multline}\label{PavlenkoVA11}
H_2=-\frac{\lambda_1^2\lambda_2(\lambda_1-1)^2}{\lambda_1-\lambda_2}\mu_1^2+
\frac{\lambda_1\lambda_2^2(\lambda_2-1)^2}{\lambda_1-\lambda_2}\mu_2^2+\\+
\frac{\lambda_1^2\lambda_2(\lambda_1-1)^2}{\lambda_1-\lambda_2}\left(\frac{\kappa_0-1}{\lambda_1}-
\frac{\gamma_1\tau_2}{\lambda_1^2}+\frac{\kappa_1}{\lambda_1-1}-\frac{\gamma_2\tau_1}{(\lambda_1-1)^2}\right)\mu_1-\\-
\frac{\lambda_1\lambda_2^2(\lambda_2-1)^2}{\lambda_1-\lambda_2}\left(\frac{\kappa_0-1}{\lambda_2}-
\frac{\gamma_1\tau_2}{\lambda_2^2}+\frac{\kappa_1}{\lambda_2-1}-\frac{\gamma_2\tau_1}{(\lambda_2-1)^2}\right)\mu_2-\kappa\lambda_1\lambda_2
\end{multline}
Во второй форме соответствующие гамильтонианы полиномиальны по координатам. Система $H^{2+2+1}$ в этом случае имеет вид:
\begin{equation}\label{PavlenkoVA12}
\frac{\partial q_k}{\partial s_j}=\frac{\partial H_j}{\partial p_k},\quad 
\frac{\partial p_k}{\partial s_j}=-\frac{\partial H_j}{\partial q_k}\quad (j,k=1,2)
\end{equation}
соответствующие гамильтонианы $H_i(s_1,s_2,q_1,q_2,p_1,p_2)$ в этом случае задаются формулами:
\begin{multline}\label{PavlenkoVA13}
s_1^2H_1=q_1^2(q_1-s_1)p_1^2+2q_1^2q_2p_1p_2+q_1q_2^2p_2^2-\\-((\kappa_0-1)q_1^2+\kappa_1q_1(q_1-s_1)+\gamma_2(q_1-s_1)+\gamma_2s_1q_2)p_1-\\-
((\kappa_0+\kappa_1-1)q_1q_2+\gamma_1s_2q_1+\gamma_2q_2)p_2+\kappa q_1,
\end{multline}
\begin{multline}\label{PavlenkoVA14}
-s_2H_2=q_1^2q_2p_1^2+2q_1q_2^2p_1p_2+q_2^2(q_2-1)p_2^2-((\kappa_0+\kappa_1-1)q_1q_2+\gamma_1s_2q_1+\gamma_2q_2)p_1-\\-
\left((\kappa_0-1)q_2(q_2-1)+\kappa_1q_2^2+\frac{\gamma_1s_2}{s_1}q_1+\gamma_1 s_2(q_2-1)\right)p_2+\kappa q_2.
\end{multline}

В статье~[16] указано соответствующее симплектическое преобразование, которое связывает упомянутые выше две формы
\begin{equation}\label{PavlenkoVA15}
q_1=\frac{(\lambda_1-1)(\lambda_2-1)}{\tau_1}, \quad q_2=\lambda_1\lambda_2,\quad s_1=\frac{1}{\tau_1},\quad s_2=-\tau_2,
\end{equation}

В статье~[17] приведена еще одна форма гамильтоновой системы $H^{2+2+1}$, у которой соответствующие гамильтонианы 
также полиномиальны по координатам
\begin{equation}\label{PavlenkoVA16}
\frac{\partial Q_k}{\partial t_j}=\frac{\partial K_j}{\partial P_k},\quad 
\frac{\partial P_k}{\partial t_j}=-\frac{\partial K_j}{\partial Q_k}\quad (j,k=1,2),
\end{equation}
где гамильтонианы $K_i(t_1,t_2,Q_1,Q_2,P_1,P_2)$ имеют вид:
\begin{multline}\label{PavlenkoVA17}
t_1K_1=P_1^2Q_1(Q_1-1)^2+[(\theta^1+\theta_1^\infty)(Q_1-1)+(\theta_2^\infty-\theta^1)Q_1(Q_1-1)+t_1Q_1]P_1-\\-
\theta^1\theta_2^\infty(Q_1-1)+(P_1Q_1^2-\theta^1Q_1-P_1)P_2Q_2+
P_1Q_2-\frac{t_2}{t_1}(P_1Q_1-P_1-\theta^1)(P_2Q_1-P_2+1),
\end{multline}
\begin{multline}\label{PavlenkoVA18}
t_2K_2=P_2^2Q_2^2-P_2Q_2^2-\theta^0P_2Q_2+t_2P_2-\theta_2^\infty Q_2-P_1Q_1Q_2+\\+\frac{t_2}{t_1}(P_1Q_1-P_1-\theta^1)(P_2Q_1-P_2+1),
\end{multline}
где постоянные $\theta^0$, $\theta^1$, $\theta_1^\infty$, $\theta_2^\infty$ удовлетворяют условию Фукса-Хукухары: 
\[
\theta^0+\theta^1+\theta_1^\infty+\theta_2^\infty=0
\]

В этой статье также сказано, что на решениях уравнений \eqref{PavlenkoVA16} с гамильтонианами \eqref{PavlenkoVA17}, \eqref{PavlenkoVA18} совместна 
следующая система линейных ОДУ
\begin{equation}\label{PavlenkoVA19}
\begin{cases}
\frac{\partial Y}{\partial\eta}=\left(\frac{A_0^{(-1)}}{\eta^2}+\frac{A_0^{(0)}}{\eta}+\frac{A_1^{(0)}}{\eta-1}+A_\infty\right)Y\\
\frac{\partial Y}{\partial t_1}=(E_2\eta+B_1+\frac{A_0^{(-1)}}{t_1\eta})Y\\
\frac{\partial Y}{\partial t_2}=-\frac{A_0^{(-1)}}{t_2\eta}Y
\end{cases}
\end{equation}
с матричными коэффициентами
\[
A_0^{(-1)}=\frac{t_2}{t_1}
\begin{pmatrix}
1-P_2& uP_2\\
\frac{1-P_2}{u}& P_2
\end{pmatrix}, 
\]
\[
A_0^{(0)}=
\begin{pmatrix}
P_1Q_1-\theta^1-\theta_1^\infty& -u(P_1Q_1+P_2Q_2+\theta_2^\infty) \\
\frac{P_1Q_1+(1-P_2)Q_2-\theta^1-\theta_1^\infty}{u}& -P_1Q_1-\theta_2^\infty
\end{pmatrix},
\]
\[
A_1^{(0)}=
\begin{pmatrix}
-P_1Q_1+\theta^1& uP_1 \\
\frac{\theta^1Q_1-P_1Q_1^2}{u}& P_1Q_1
\end{pmatrix}, 
\]
\[
A_\infty=
\begin{pmatrix}
0& 0 \\ 0& t_1
\end{pmatrix}, \quad
E_2=
\begin{pmatrix}
0& 0 \\ 0& 1
\end{pmatrix},
\]
\[
B_1=\frac{1}{t_1}
\begin{pmatrix}
0& (A_0^{(0)})_{12}+(A_1^{(0)})_{12} \\ (A_0^{(0)})_{21}+(A_1^{(0)})_{21} & 0
\end{pmatrix},
\]
зависящими также от совместного решения следующих линейных ОДУ:
\[
t_1u_{t_1}=u(\theta^1(1-Q_1)+P_1(1-Q_1)^2+\theta_1^\infty-\theta_2^\infty),\quad t_2u_{t_2}=-uQ_2.
\]
Легко видеть, что замена 
\[
Y=\exp{\left(\frac{\eta t_1}{2}-\frac{t_2}{2\eta t_1}+\frac{\theta^0}{2}\ln|\eta|+\frac{\theta^1}{2}\ln|\eta-1|\right)}Z
\]
совместные системы ИДМ \eqref{PavlenkoVA19} переводит в эквивалентные им совместные системы
\begin{equation}\label{PavlenkoVA20}
\begin{cases}
\frac{\partial Z}{\partial\eta}=\left(\frac{B_0^{(-1)}}{\eta^2}+\frac{B_0^{(0)}}{\eta}+\frac{B_1^{(0)}}{\eta-1}+B_\infty\right)Z\\
\frac{\partial Z}{\partial t_1}=(F_2\eta+B_1+\frac{B_0^{(-1)}}{t_1\eta})Z\\
\frac{\partial Z}{\partial t_2}=-\frac{B_0^{(-1)}}{t_2\eta}Z
\end{cases}
\end{equation}
с матричными коэффициентами
\[
B_0^{(-1)}=\frac{t_2}{t_1}
\begin{pmatrix}
0.5-P_2& uP_2\\
\frac{1-P_2}{u}& P_2-0.5
\end{pmatrix}, 
\]
\[
B_0^{(0)}=
\begin{pmatrix}
P_1Q_1+0.5\theta^0+\theta_2^\infty& -u(P_1Q_1+P_2Q_2+\theta_2^\infty) \\
\frac{P_1Q_1+(1-P_2)Q_2-\theta^1-\theta_1^\infty}{u}& -P_1Q_1-0.5\theta^0-\theta_2^\infty
\end{pmatrix},
\]
\[
B^{(0)}_1=
\begin{pmatrix}
-P_1Q_1+0.5\theta^1& uP_1 \\
\frac{\theta^1Q_1-P_1Q_1^2}{u}& P_1Q_1-0.5\theta^1
\end{pmatrix}, 
\]
\[
B_\infty=
\begin{pmatrix}
-\frac{t_1}{2}& 0 \\ 0& \frac{t_1}{2}
\end{pmatrix}, \quad
F_2=
\begin{pmatrix}
-0.5& 0 \\ 0& 0.5
\end{pmatrix},
\]
имеющими нулевой след. В \eqref{PavlenkoVA20} выполним следующую замену: 
\[
\tau_1=t_1,\quad \tau_2=\frac{t_2}{t_1}.
\]
Получим следующую совместную систему: 
\begin{equation}\label{PavlenkoVA21}
\begin{cases}
\frac{\partial Z}{\partial\eta}=\left(\frac{B_0^{(-1)}}{\eta^2}+\frac{B_0^{(0)}}{\eta}+\frac{B_1^{(0)}}{\eta-1}+B_\infty\right)Z\\
\frac{\partial Z}{\partial \tau_1}=(F_2\eta+B_1)Z\\
\frac{\partial Z}{\partial \tau_2}=-\frac{B_0^{(-1)}}{\tau_2\eta}Z
\end{cases}
\end{equation}
Матричные коэффициенты, которые зависят от времени, перепишутся в виде: 
\begin{multline*}
B_0^{(-1)}=\tau_2
\begin{pmatrix}
0.5-P_2& uP_2\\
\frac{1-P_2}{u}& P_2-0.5
\end{pmatrix}, \quad 
B_\infty=
\begin{pmatrix}
-\frac{\tau_1}{2}& 0 \\ 0& \frac{\tau_1}{2}
\end{pmatrix}, \\
B_1=\frac{1}{\tau_1}
\begin{pmatrix}
0& (A_0^{(0)})_{12}+(A_1^{(0)})_{12} \\ (A_0^{(0)})_{21}+(A_1^{(0)})_{21} & 0
\end{pmatrix}
\end{multline*}

Используем приведенную выше матричную форму уравнений ИДМ гамильтоновой системы $H^{2+2+1}$ 
для построения решений соответствующих эволюционных уравнений. Справедливо следующее 

\textbf{Утверждение.} {\it Два последних  уравнения системы \eqref{PavlenkoVA21} есть условие совместности нелинейной интегрируемой системы, 
котрая является неоднородным обобщением известной комплексифицированной системы Полмайера -- Редже -- Лунда -- Гетманова. }

{\bf Доказательство.} Обозначим: 
\begin{equation}\label{PavlenkoVA22}
U=F_2\eta+B_1,\quad V=-\frac{B_0^{(-1)}}{\tau_2\eta}
\end{equation}
Матрицы $U$ и $V$ перепишем в виде:
\begin{equation}\label{PavlenkoVA23}
U=F_2\eta+\frac{1}{\tau_1}
\begin{pmatrix}
0& b \\ a& 0
\end{pmatrix},\quad 
V=\frac 1 \eta 
\begin{pmatrix}
c& d \\ e& -c
\end{pmatrix},
\end{equation}
где 
\begin{multline}\label{PavlenkoVA24}
a=(A_0^{(0)})_{21}+(A_1^{(0)})_{21},\quad b=(A_0^{(0)})_{12}+(A_1^{(0)})_{12},\\ c=P_2-0.5,\quad d=-uP_2, 
\quad e=\frac{P_2-1}{u}.
\end{multline}
Условием совместности последних двух уравнений системы \eqref{PavlenkoVA21} является равенство
\begin{equation}\label{PavlenkoVA25}
U_{\tau_2}-V_{\tau_1}+[U,V]=0
\end{equation}
Справедливость \eqref{PavlenkoVA25} с учетом обозначений \eqref{PavlenkoVA22}, \eqref{PavlenkoVA23}, \eqref{PavlenkoVA24} означает, что 
имеет место замкнутая система дифференциальных уравнений
\begin{equation}\label{PavlenkoVA26}
\tau_1c_{\tau_1}=eb-ad,\quad \tau_1d_{\tau_1}=-2bc,\quad \tau_1e_{\tau_1}=2ac,
\quad b_{\tau_2}=\tau_1 d,\quad a_{\tau_2}=-e\tau_1.
\end{equation}
Из \eqref{PavlenkoVA24} следует, что имеет место равенство 
\begin{equation}\label{PavlenkoVA27}
c^2+de=\frac{1}{4}.
\end{equation}
С учетом \eqref{PavlenkoVA27} из \eqref{PavlenkoVA26} вытекает справедливость формул 
\[
\tau_1 b_{\tau_1\tau_2}=b_{\tau_2} -b\sqrt{\tau_1^2+4a_{\tau_2}b_{\tau_2}},\quad 
\tau_1 a_{\tau_1\tau_2}=a_{\tau_2}-a\sqrt{\tau_1^2+4a_{\tau_2}b_{\tau_2}}, 
\]
 а, это означает, что пара $a$ и $b$ удовлетворяет неоднородному обобщению комплексифицированной системе 
Полмайера -- Редже -- Лунда -- Гетманова (см. работы~[29, 30]). Таким образом, утверждение доказано. 

\textbf{2. Построение решений аналогов временных уравнения Шредингера.} Основным результатом настоящей работы является следующая

\textbf{Теорема.} {\it Существуют решения уравнений \eqref{PavlenkoVA7}, соответствующих гамильтоновой системе 
$H^{2+2+1}$ с рациональными коэффициентами, явным образом выражаются в терминах решений системы ОДУ \eqref{PavlenkoVA21}.}

{\bf Доказательство.} $2\times 2$ матрица 
\begin{equation*}
M=Z^{-1}(\tau_1,\tau_2,\eta)Z(\tau_1,\tau_2,\zeta),
\end{equation*}
образованная по совместному фундаментальному решению $Z$ линейных cистем \eqref{PavlenkoVA21},
удовлетворяет двум следующим {\it скалярным} эволюционным уравнениям с временами $\tau_1$ и $\tau_2$: 
\begin{multline}\label{PavlenkoVA28}
\tau_1M_{\tau_1}=\frac{\zeta^2(\zeta-1)}{\zeta-\eta}M_{\zeta\zeta}-\frac{\eta^2(\eta-1)}{\zeta-\eta}M_{\eta\eta}+\\+
\frac{\zeta(\zeta^2-3\zeta\eta+2\eta)}{(\zeta-\eta)^2}M_\zeta+\frac{\eta(\eta^2-3\zeta\eta+2\zeta)}{(\zeta-\eta)^2}M_\eta+g_1M,
\end{multline}
\begin{equation}\label{PavlenkoVA29}
\tau_2M_{\tau_2}=\frac{\zeta^2\eta(\zeta-1)}{\zeta-\eta}M_{\zeta\zeta}-\frac{\zeta\eta^2(\eta-1)}{\zeta-\eta}M_{\eta\eta}+
\frac{\zeta\eta(\zeta+\eta-2\zeta\eta)}{(\zeta-\eta)^2}(M_\zeta+M_\eta)+g_2M.
\end{equation}
Здесь функции $g_1(\tau_1,\tau_2,\zeta,\eta,P_1,P_2,Q_1,Q_2)$ и $g_2(\tau_1,\tau_2,\zeta,\eta,P_1,P_2,Q_1,Q_2)$ задаются формулами 
\begin{multline*}
g_1(\tau_1,\tau_2,\zeta,\eta,P_1,P_2,Q_1,Q_2)=\frac{\tau_2^2(\zeta\eta-\zeta-\eta)}{4\zeta^2\eta^2}-\frac{\theta^0\tau_2}{2\zeta\eta}
+\tau_1\tau_2(0,5-P_2)+\det(B^{(0)}_0)\\+\frac{(\theta^1)^2(\zeta+\eta-\zeta\eta)}{4(\zeta-1)(\eta-1)}-2(B_0^{(0)})_{11}(B_1^{(0)})_{11}-
(B_0^{(0)})_{21}(B_1^{(0)})_{12}-(B_0^{(0)})_{12}(B_1^{(0)})_{21}-\\-\tau_1(P_1Q_1+0.5\theta^0+\theta_2^\infty)+
0.5\tau_1(\theta_2^\infty-\theta_1^\infty)(\zeta+\eta)+0.25\tau_1^2(\zeta+\eta-\zeta^2-\eta^2-\zeta\eta),
\end{multline*}
\begin{multline*}
g_2(\tau_1,\tau_2,\zeta,\eta,P_1,P_2,Q_1,Q_2)=\frac{\tau_2^2(\zeta\eta(\zeta+\eta)-\zeta^2-\eta^2-\zeta\eta)}{4\zeta^2\eta^2}+
\frac{\theta^0\tau_2(\zeta\eta-\zeta-\eta)}{2\zeta\eta}+\\+\tau_1\tau_2(0.5-P_2)+\det(B^{(0)}_0)+
\frac{(\theta^1)^2\zeta\eta}{4(\zeta-1)(\eta-1)}+2(B_0^{(-1)})_{11}(B_1^{(0)})_{11}+\\+(B_0^{(-1)})_{21}(B_1^{(0)})_{12}+
(B_0^{(-1)})_{12}(B_1^{(0)})_{21}+0.5\tau_1(\theta_2^\infty-\theta_1^\infty)\zeta\eta+0.25\tau_1^2\zeta\eta(1-\zeta-\eta).
\end{multline*}
Сделаем замену: 
\[
M=\exp{(S(\tau_1,\tau_2))}W,
\]
где функция $S$ удовлетворяет непротиворечивым равенствам
\begin{multline*}
\tau_1S_{\tau_1}=\tau_1\tau_2(0,5-P_2)+\det(B^{(0)}_0)-\\-2(B_0^{(0)})_{11}(B_1^{(0)})_{11}-
(B_0^{(0)})_{21}(B_1^{(0)})_{12}-(B_0^{(0)})_{12}(B_1^{(0)})_{21}-\tau_1P_1Q_1,
\end{multline*}
\begin{multline*}
\tau_2S_{\tau_2}=\tau_1\tau_2(0.5-P_2)+\det(B^{(0)}_0)+\\+2(B_0^{(-1)})_{11}(B_1^{(0)})_{11}+(B_0^{(-1)})_{21}(B_1^{(0)})_{12}+
(B_0^{(-1)})_{12}(B_1^{(0)})_{21}.
\end{multline*}
Получим, что уравнения \eqref{PavlenkoVA28}, \eqref{PavlenkoVA29} сводятся к следующим двум уравнениям
\begin{multline}\label{PavlenkoVA30}
\tau_1W_{\tau_1}=\frac{\zeta^2(\zeta-1)}{\zeta-\eta}W_{\zeta\zeta}-\frac{\eta^2(\eta-1)}{\zeta-\eta}W_{\eta\eta}+\\+
\frac{\zeta(\zeta^2-3\zeta\eta+2\eta)}{(\zeta-\eta)^2}W_\zeta+\frac{\eta(\eta^2-3\zeta\eta+2\zeta)}{(\zeta-\eta)^2}W_\eta+g_3W,
\end{multline}
\begin{equation}\label{PavlenkoVA31}
\tau_2W_{\tau_2}=\frac{\zeta^2\eta(\zeta-1)}{\zeta-\eta}W_{\zeta\zeta}-\frac{\zeta\eta^2(\eta-1)}{\zeta-\eta}W_{\eta\eta}+
\frac{\zeta\eta(\zeta+\eta-2\zeta\eta)}{(\zeta-\eta)^2}(W_\zeta+W_\eta)+g_4W.
\end{equation}
Здесь функции $g_3(\tau_1,\tau_2,\zeta,\eta)$ и $g_4(\tau_1,\tau_2,\zeta,\eta)$ уже не зависят от переменных 
$P_i$, $Q_i$ $(i=1,2)$ и имеют следующий вид:
\begin{multline*}
g_3(\tau_1,\tau_2,\zeta,\eta)=\frac{\tau_2^2(\zeta\eta-\zeta-\eta)}{4\zeta^2\eta^2}-\frac{\theta^0\tau_2}{2\zeta\eta}
+\frac{(\theta^1)^2(\zeta+\eta-\zeta\eta)}{4(\zeta-1)(\eta-1)}-\tau_1(0.5\theta^0+\theta_2^\infty)+\\+
0.5\tau_1(\theta_2^\infty-\theta_1^\infty)(\zeta+\eta)+0.25\tau_1^2(\zeta+\eta-\zeta^2-\eta^2-\zeta\eta),
\end{multline*}
\begin{multline*}
g_4(\tau_1,\tau_2,\zeta,\eta)=\frac{\tau_2^2(\zeta\eta(\zeta+\eta)-\zeta^2-\eta^2-\zeta\eta)}{4\zeta^2\eta^2}+
\frac{\theta^0\tau_2(\zeta\eta-\zeta-\eta)}{2\zeta\eta}+\\+\frac{(\theta^1)^2\zeta\eta}{4(\zeta-1)(\eta-1)}+
0.5\tau_1(\theta_2^\infty-\theta_1^\infty)\zeta\eta+0.25\tau_1^2\zeta\eta(1-\zeta-\eta).
\end{multline*}
Далее выполним замену переменных: 
\[
x=\frac{\zeta}{\zeta-1},\quad y=\frac{\eta}{\eta-1}, 
\]
которая сводит уравнения \eqref{PavlenkoVA30}, \eqref{PavlenkoVA31} к уравнениям: 
\begin{multline}\label{PavlenkoVA32}
\tau_1W_{\tau_1}=-\frac{x^2(x-1)^2(y-1)}{x-y}W_{xx}+\frac{y^2(y-1)^2(x-1)}{x-y}W_{yy}+\\+
\frac{x(x-1)(y-1)(x^2+xy-2y)}{(x-y)^2}W_x+\frac{y(y-1)(x-1)(y^2+xy-2x)}{(x-y)^2}W_y+g_5W,
\end{multline}
\begin{multline}\label{PavlenkoVA33}
\tau_2W_{\tau_2}=-\frac{x^2(x-1)^2y}{x-y}W_{xx}+\frac{y^2(y-1)^2x}{x-y}W_{yy}+\\+
\frac{xy(x+y)(x-1)^2}{(x-y)^2}W_x+\frac{xy(x+y)(y-1)^2}{(x-y)^2}W_y+g_6W,
\end{multline}
где функции $g_5(\tau_1,\tau_2,x,y)$ и $g_6(\tau_1,\tau_2,x,y)$ принимают вид:
\begin{multline*} 
g_5(\tau_1,\tau_2,x,y)=\frac{\tau_2^2(x+y-xy)(x-1)(y-1)}{4x^2y^2}-\frac{\theta^0\tau_2(x-1)(y-1)}{2xy}+\\+
0.25(\theta^1)^2(xy-x-y)-\tau_1(0.5\theta^0+\theta_2^\infty)+\frac{\tau_1(\theta_2^\infty-\theta_1^\infty)(2xy-x-y)}{2(x-1)(y-1)}-\\-
\frac{\tau_1^2(x^2y^2-3xy+x+y)}{4(x-1)^2(y-1)^2},
\end{multline*}
\begin{multline*}
g_6(\tau_1,\tau_2,x,y)=\frac{\tau_2^2(2x^2y+2xy^2-x^2y^2-xy-x^2-y^2)}{4x^2y^2}+
\frac{\theta^0\tau_2(x+y-xy)}{2xy}+\\+0.25(\theta^1)^2xy+
\frac{\tau_1(\theta_2^\infty-\theta_1^\infty)xy}{2(x-1)(y-1)}+\frac{\tau_1^2xy(1-xy)}{4(x-1)^2(y-1)^2}.
\end{multline*}
Наконец, в уравнениях \eqref{PavlenkoVA32}, \eqref{PavlenkoVA33} сделаем замену: 
\[
W=e^{f_1(x,y,\tau_1,\tau_2)+f_2(\tau_1,\tau_2)}\Psi,
\]
где 
\begin{multline*}
f_1(x,y,\tau_1,\tau_2)=(0.5\kappa_0-1)\ln(|xy|)+0.5\kappa_1(\ln|(x-1)(y-1)|)+\ln|x-y|+\\+
\frac{\gamma_1\tau_2(x+y)}{2xy}+\frac{\gamma_2\tau_1(x+y-2)}{2(x-1)(y-1)}.
\end{multline*}
\begin{multline*}
f_2(\tau_1,\tau_2)=\frac{((\kappa_1-2)^2-(\theta^1)^2-4)\ln\tau_1}{4}+\frac{(\kappa_0-2)^2\ln\tau_2}{4}-\\-
\frac{((\kappa_0-2)\gamma_2+\theta^0+2\theta_2^\infty)\tau_1}{2}+
\frac{((\kappa_1-2)\gamma_1+\theta^0)\tau_2}{2}+\frac{\gamma_1\gamma_2\tau_1\tau_2}{2}.
\end{multline*}
При этом положим, что:
\begin{multline*}
\kappa=\frac{(\kappa_0-2)^2}{4}+\frac{(\kappa_1-2)^2}{4}+\frac{\kappa_0\kappa_1}{2}-\frac{(\theta^1)^2}{4}-2, \quad
\theta^0=(\kappa_0-2)\gamma_1, \\ \theta_2^\infty-\theta_1^\infty=(\kappa_1-2)\gamma_2.
\end{multline*}
Получим следующие уравнения:
\begin{multline}\label{PavlenkoVA34}
\tau_1\Psi_{\tau_1}=-\frac{x^2(x-1)^2(y-1)}{x-y}\Psi_{xx}+\frac{y^2(y-1)^2(x-1)}{x-y}\Psi_{yy}-\\-
\frac{x^2(x-1)^2(y-1)}{(x-y)}\left(\frac{\kappa_0}{x}-\frac{\gamma_1\tau_2}{x^2}+\frac{\kappa_1-1}{x-1}-
\frac{\gamma_2\tau_1}{(x-1)^2}\right)\Psi_x+\\+\frac{(x-1)y^2(y-1)^2}{(x-y)}\left(\frac{\kappa_0}{y}-
\frac{\gamma_1\tau_2}{y^2}+\frac{\kappa_1-1}{y-1}-\frac{\gamma_2\tau_1}{(y-1)^2}\right)\Psi_y+g_7\Psi,
\end{multline}
\begin{multline}\label{PavlenkoVA35}
\tau_2\Psi_{\tau_2}=-\frac{x^2(x-1)^2y}{x-y}\Psi_{xx}+\frac{y^2(y-1)^2x}{x-y}\Psi_{yy}-\\-
\frac{x^2(x-1)^2y}{(x-y)}\left(\frac{\kappa_0-1}{x}-\frac{\gamma_1\tau_2}{x^2}+\frac{\kappa_1}{x-1}-
\frac{\gamma_2\tau_1}{(x-1)^2}\right)\Psi_x+\\+\frac{xy^2(y-1)^2}{(x-y)}\left(\frac{\kappa_0-1}{y}-
\frac{\gamma_1\tau_2}{y^2}+\frac{\kappa_1}{y-1}-\frac{\gamma_2\tau_1}{(y-1)^2}\right)\Psi_y+g_8\Psi,
\end{multline}
где функции $g_7(\tau_1,\tau_2,x,y)$ и $g_8(\tau_1,\tau_2,x,y)$ имеют вид:
\begin{multline*} 
g_7(\tau_1,\tau_2,x,y)=-\kappa(x-1)(y-1)+\frac{(\gamma_1^2-1)\tau_2^2(x-1)(y-1)(x+y-xy)}{4x^2y^2}+\\+
\frac{(\gamma_2^2-1)\tau_1^2(x^2y^2-3xy+x+y)}{4(x-1)^2(y-1)^2}+\frac{4(x-1)(y-1)xy}{(x-y)^2},
\end{multline*}
\begin{multline*}
g_8(\tau_1,\tau_2,x,y)=-\kappa xy+
\frac{(\gamma_1^2-1)\tau_2^2(x^2y^2+xy+x^2+y^2-2x^2y-2xy^2)}{4x^2y^2}+\\+
\frac{(\gamma_2^2-1)\tau_1^2xy(xy-1)}{4(x-1)^2(y-1)^2}+\frac{2xy(2xy-x-y)}{(x-y)^2}.
\end{multline*}
Далее, за счет справедливости коммутационных соотношений  Гейзенберга
\[
\frac{\partial}{\partial x}x-x\frac{\partial}{\partial x}=1,\quad 
\frac{\partial}{\partial y}y-y\frac{\partial}{\partial y}=1
\]
уравнения \eqref{PavlenkoVA34}, \eqref{PavlenkoVA35} символически записываются в виде \eqref{PavlenkoVA7},  
определяемых гамильтонианами \eqref{PavlenkoVA10}, \eqref{PavlenkoVA11} с рациональными коэффициентами 
гамильтоновой системы \eqref{PavlenkoVA9}. Тем самым, основная теорема о построении решений 
аналогов временных уравнения Шредингера доказана. 

Справедливо также следующее 

\textbf{Следствие.} {\it Существуют решения уравнений \eqref{PavlenkoVA8}, соответствующих гамильтоновой системе 
$H^{2+2+1}$ с полиномиальными коэффициентами, явным образом выражаются в терминах решений системы ОДУ \eqref{PavlenkoVA21}. }

{\bf Доказательство.} Легко проверить, что квантовый аналог замены \eqref{PavlenkoVA15} 
\begin{equation*}
r=\frac{(x-1)(y-1)}{\tau_1},\quad\rho=xy,\quad s_1=\frac{1}{\tau_1},\quad s_2=-\tau_2,
\end{equation*}
уравнения \eqref{PavlenkoVA34}, \eqref{PavlenkoVA35} сводит к эволюционным уравнениям 
\begin{multline*}
s_1^2\Psi_{s_1}=r^2(r-s_1)\Psi_{rr}+2r^2\rho\Psi_{r\rho}+r\rho^2\Psi_{\rho\rho}+
((\kappa_0-1)r^2+(\kappa_1 r+\gamma_2)(r-s_1)+\gamma_2 s_1\rho)\Psi_r+\\+
((\kappa_0+\kappa_1-1)r\rho+\gamma_1 s_2 r+\gamma_2\rho)\Psi_\rho+\kappa r\Psi,
\end{multline*}
\begin{multline*}
-s_2\Psi_{s_2}=r^2\rho\Psi_{rr}+2r\rho^2\Psi_{r\rho}+\rho^2(\rho-1)\Psi_{\rho\rho}+
((\kappa_0+\kappa_1-1)r\rho+\gamma_1 s_2 r+\gamma_2\rho)\Psi_r+\\+
\left((\kappa_0-1)\rho(\rho-1)+\kappa_1\rho^2+\frac{\gamma_1 s_2 r}{s_1}+\gamma_1 s_2(\rho-1)\right)\Psi_\rho+\kappa\rho\Psi, 
\end{multline*}
которые 
за счет справедливости соотношений Гейзенберга
\[
\frac{\partial}{\partial r}r-r\frac{\partial}{\partial r}=1,\quad 
\frac{\partial}{\partial\rho}\rho-\rho\frac{\partial}{\partial\rho}=1, 
\]
символически могут быть записаны в виде \eqref{PavlenkoVA8}, определяемых полиномиальными гамильтонианами \eqref{PavlenkoVA13}, \eqref{PavlenkoVA14} 
с полиномиальными коэффициентами гамильтоновой системы \eqref{PavlenkoVA12}. Следствие доказано. 

\smallskip

\textbf{Заключение.} Таким образом, на сегодняшний день в иерархии Кимуры остался всего лишь один представитель $H^{3+1+1}$ для которого 
аналоги временных уравнений Шредингера ещё не построены. 

Автор заявляет, что стороны не имеют конфликта интересов. 

\bigskip

\begin{center}{СПИСОК ЛИТЕРАТУРЫ}\end{center}{\small
\lit{1}{Cулейманов~Б.И.}{Гамильтонова структура уравнений Пенлеве
и метод изомонодромных деформаций //~ Асимптотические свойства решений дифференциальных уравнений. Институт математики. Уфа. 1988, С.~93 -- 102.}
\lit{2}{Cулейманов~Б.И.}{Гамильтоновocть уравнений Пенлеве и метод изомонодромных деформаций //  Дифференциальные уравнения. 1994. T.~30. №~5. С.~791 -- 796. }
\lit{3}{Garnier~R.}{Sur des equations diff\'{e}rentielles du troisieme ordre 
dont l'integrale generale est uniforme et sur une classe d'equations nouvelles d'ordre 
superieur dont l'integrale generale a ses points critiques fixes // Ann. Sci. Ecole Normale Sup. 1912. T.29. №~3. P.~1 -- 126. }
\lit{4}{Bloemendal~A., Virag~B.}{Limits of spiked random matrices $II$ // Ann. Probab. 2016. T.~44. №~4. P.~2726 -- 2769.}
\lit{5}{Conte~R.}{Generalized Bonnet surfaces and Lax pairs of PVI // J. Math. Phys. 2017. T.~58. №~10. P.~1 -- 31}
\lit{6}{Grundland~A.M., Riglioni~D.} {\it Classical-quantum correspondence for shape-invariant systems // 
J. Phys. A. 2015. T.~48. №~24. P.~245201 -- 245215. }
\lit{7}{Levin~A.M., Olshanetsky~M.A., Zotov~A.V.}{Planck Constant as Spectral Parameter in Integrable Systems and KZB Equations // 
Journal of High Energy Physics. 2014. T.~10. P.~1--29. DOI: 10.1007/JHEP10(2014)109.}
\lit{8}{Nagoya~H.}{Hypergeometric solutions to Schr\"odinger equation for the quantum
Painlev\'{e} equations. // J. Math. Phys. 2011. T.~52. №~8. P.~1 -- 16.}
\lit{9}{Rosengren~H.}{Special polynomials related to the supersymmetric 
eight-vertex model: a summary. // Commun. Math. Phys. 2015. T.~15. №~3. P.~1143 -- 1170.}
\lit{10}{Rumanov~I.}{Painlev\'e representation of Tracy-Widom$_\beta$ distribution for $\beta = 6$. // 
Comm. Math. Phys. 2016. T.~342. №~3. P.~843 -- 868.}
\lit{11}{Zabrodin~A., Zotov~A.}{Quantum Painlev\'{e}-Calogero correspondence. // J. Math. Phys. 2012. T.~53. №~7. P.~1 -- 19.}
\lit{12}{Zabrodin~A., Zotov~A.}{Classical-quantum correspondence and functional 
relations for Painlev\'{e} equations. //  Constr. Approx. 2015. T.~41. №~3. P.~385 -- 423.}
\lit{13}{Левин~А.М., Ольшанецкий~М.А., Зотов~А.В.}{О некоторых асимптотических свойствах решений уравнений типа Эмдена--Фаулера // 
Дифференц. уравнения. 1981. Т.~17. №~4. С.~749 -- 750.}
\lit{14}{Новиков~Д.П.}{О системе Шлезингера с матрицами размера $2\times2$ и уравнении Белавина -- Полякова -- Замолодчикова. // 
ТМФ. 2009. T.~161. №~2. P.~191 -- 203.}
\lit{15}{Сулейманов~Б.И.}{Квантовые аспекты интегрируемости третьего уравнения 
Пенлеве и решения временного уравнения Шредингера с потенциалом Морса. //   
Уфимский математический журнал. 2016. T.~8. №~3. P.~141 -- 159.}
\lit{16}{Kimura~H.}{The degeneration of the two dimensional Garnier system and the polynomial Hamiltonian structure. // Annali 
di Matematica pura et applicata IV. V.~155. №~1. P.~25 -- 74.}
\lit{17}{Kawakami~H., Nakamura~A., Sakai~H.}{Degeneration scheme of 4-dimensional Painleve-type equations. // arXiv:1209.3836. 2012.}
\lit{18}{Sakai~H.}{Isomonodromic deformation and 4-dimensional Painleve-type equations. // Tech. Report, Univ. Tokyo. Tokyo. 2010.}
\lit{19}{Kawakami~H., Nakamura~A., Sakai~H.}{Toward a classification of 4-dimensional Painlev?e-type equations. // Contemporary Mathematics, 593,
eds. A. Dzhamay, K. Maruno, V. U. Pierce, AMS, Providence, RI. 2013. P.~143 -- 161.}
\lit{20}{Kawamuko~H.}{On qualitative properties and asymptotic behavior of solutions to higher-order nonlinear differential equations // 
WSEAS Transact. on Math. 2017. V.~16. №~5. P.~39 -- 47.}
\lit{21}{Цегельник~В.В.}{Некоторые аналитические свойства и приложения решений уравнений Пенлеве-типа. // Минск. 2007. С.~1 -- 223.}
\lit{22}{Цегельник~В.В.}{О свойствах решений двух дифференциальных уравнений второго порядка со свойством Пенлеве. //  
ТМФ. 2021. T.~206. №~3. C.~361 -- 367.}
\lit{23}{Сулейманов~Б.И.}{<<Квантования>>  высших гамильтоновых аналогов уравнений Пенлеве I и II с двумя степенями свободы. // 
Функциональный анализ и его приложения. 2014. T.~48. Выпуск~3. С.~52 -- 62.}

\pagebreak

\lit{24}{Новиков~Д.П., Сулейманов~Б.И.}{<<Квантования>> изомонодромной 
гамильтоновой системы Гарнье с двумя степенями свободы. ТМФ. 2016. T.~187. №~1. C.~39 -- 57.}
\lit{25}{Павленко~В.А., Сулейманов~Б.И.}{Решения аналогов временных уравнений Шредингера, 
определяемых изомонодромной гамильтоновой системой $H^{2+1+1+1}$. // Уфимский математический журнал. 2018. Т.~10. №~4. С.~92 -- 102} 
\lit{26}{Павленко~В.А., Сулейманов~Б.И.}{Решения аналогов временных уравнений Шредингера, 
определяемых изомонодромной гамильтоновой системой $H^{4+1}$. // Известия РАН, серия физическая. 2020. Т.~84. №~5. C.~695 -- 698}
\lit{27}{Павленко~В.А.}{Решения аналогов временных уравнений Шредингера, соответствующих паре гамильтоновых систем $H^{3+2}$. // 
ТМФ. 2022. T.~212. №~3. C.~340 -- 353.}
\lit{28}{Сулейманов~Б.И.}{Изомонодромное квантование  второго уравнения Пенлеве 
посредством консервативных гамильтоновых систем с двумя степенями свободы. // Алгебра и анализ. 2021. T.~33. №~6. C.~141 -- 161.}
\lit{29}{Lund~F.}{Classically solvable field theory model. // Ann. Phys. 1978. V.~115. P.~251 -- 270}
\lit{30}{Гетманов~Б.С.}{Интегрируемая модель нелинейного комплексного скалярного поля с нетривиальной 
асимптотикой солитонных решений. // ТМФ. 1979. Т.~38. №~2. С.~186 -- 194.}

} 
\newpage

СВЕДЕНИЯ ОБ АВТОРАХ (копия карточки автора для каждого автора в отдельности)\\

1.	ФИО: Павленко Виктор Александрович (Pavlenko Viktor Aleksandrovich)

2.	Дата рождения: 10.01.1986

3.	Место работы: Институт математики с вычислительным центром (Institute of Mathematics with Computing Centre)

4.	Занимаемая должность: младший научный сотрудник

5.	Ученое звание и степень: кандидат физико-математических наук

6.	Служебный адрес: Уфа, ул. Чернышевского, 112

7.	Телефон (с кодом города): +7 347 272 59 36

8.	Домашний адрес: Уфа, ул. Менделеева, 177/2, кв. 195

9.	Телефон (с кодом города): +7 917 77 318 40

10.	Электронный адрес: mail@pavlenko.school

11.	Факс: +7 347 272 59 36

12.	Основные направления научных исследований: Интегрируемость нелинейных уравнений, Уравнения Шредингера, Уравнения типа Пенлеве. 

13. Название статьи: Решения аналогов временных уравнений Шредингера, соответствующих паре гамильтоновых систем $H^{2+2+1}$. 
(Solutions to analogues of nonstationary Schrodinger equations corresponding to a pair of Hamiltonian systems $H^{2+2+1}$.)

14. УДК: 517.925

15. Раздел (рубрика), к которому относится статья:\\ Уравнения с частными производными.

Название организации указывать без сокращений, телефоны и электронный адрес указывать обязательно.

\end{document}